\documentclass[fleqn,usenatbib]{aa}
\usepackage{graphicx}
\usepackage{txfonts}

\begin{document}

   \title{Asteroseismology of the Hyades red giant and planet host $\epsilon$\,Tau\thanks{Based on observations made with the SONG telescopes operated on the Spanish Observatorio del Teide (Tenerife) and at the Chinese Delingha Observatory (Qinghai) by the Aarhus and Copenhagen Universities, by the Instituto de Astrof\'isica de Canarias and by the National Astronomical Observatories of China,
and with NASA's K2 mission.}\fnmsep\thanks{Time-series data are only available in electronic form at the CDS via anonymous ftp to cdsarc.u-strasbg.fr (130.79.128.5) or via http://cdsarc.u-strasbg.fr/viz-bin/qcat?J/A+A/622/A190}}

   \author{T. Arentoft\inst{1}\fnmsep\thanks{toar@phys.au.dk},
           F. Grundahl\inst{1},
           T. R. White\inst{1},
           D. Slumstrup\inst{1},
           R. Handberg\inst{1},
           M. N. Lund\inst{1},
           K. Brogaard\inst{1},
           M. F. Andersen\inst{1},
           V.~Silva Aguirre\inst{1},
           C. Zhang\inst{2},
           X. Chen\inst{2},
           Z. Yan\inst{3},
           B.~J.~S.~Pope\inst{4,5},
           D.~Huber\inst{6},
           H. Kjeldsen\inst{1},
           J. Christensen-Dalsgaard\inst{1},
           J. Jessen-Hansen\inst{1},
           V. Antoci\inst{1},
           S. Frandsen\inst{1},
           T.~R. Bedding\inst{7,1},
           P. L. Pall\'e\inst{8,9},
           R. A. Garcia\inst{10,11},
           L. Deng\inst{2,3},
           M. Hon\inst{12},
           D. Stello\inst{12,7,1}
            \and 
           U.~G. J{\o}rgensen\inst{13}
          }
    
   \institute{Stellar Astrophysics Centre, Dept. of Physics and Astronomy, Aarhus University, Ny Munkegade, DK-8000 Aarhus C.
              \and
              Key Laboratory of Optical Astronomy, National Astronomical Observatories, Chinese Academy of
              Sciences, Beijing, 100101, PR China
              \and 
              Department of Astronomy, China West Normal University, Nanchong 637002, PR China
              \and 
              Center for Cosmology and Particle Physics, Dept. of Physics, New York University, 726 Broadway, New York, NY 10003, USA
              \and 
              NASA Sagan Fellow
              \and 
              Institute for Astronomy, University of Hawai`i, 2680 Woodlawn Drive, Honolulu, HI 96822, USA
              \and
              Sydney Institute for Astronomy (SIfA), School of Physics, University of Sydney, NSW 2006, Australia
              \and
              Instituto de Astrof\'isica de Canarias, E-38205 La Laguna, Tenerife, Spain
              \and
              Universidad de La Laguna, Dpto. Astrof\'isica, E-38206 La Laguna, Tenerife, Spain
              \and 
              IRFU, CEA, Universit\'e Paris-Saclay, F-91191 Gif-sur-Yvette, France
              \and 
              Universit\'e Paris Diderot, AIM, Sorbonne Paris Cit\'e, CEA, CNRS, F-91191 Gif-sur-Yvette, France
              \and  
              School of Physics, The University of New South Wales, Sydney NSW 2052, Australia
              \and
              Centre for Star and Planet Formation, Niels Bohr Institute, University of Copenhagen, Øster Voldgade 5, DK-1350 Copenhagen
             }

   \date{Received 20 November 2018 / Accepted 4 January 2019}
 
  \abstract
   {Asteroseismic analysis of solar-like stars allows us to determine physical parameters such as stellar mass,
   with a higher precision compared to most other methods. Even in a well-studied 
   cluster such as the Hyades, the masses of the red giant stars are not well known, and previous mass estimates are based on model calculations (isochrones). The four known red giants in the Hyades are assumed to be clump (core-helium-burning) stars based on their positions in colour-magnitude diagrams, however asteroseismology offers an opportunity to test this assumption.}
   {Using asteroseismic techniques combined with other methods, we aim to derive physical parameters and the evolutionary stage for the planet hosting star $\epsilon$\,Tau, which is one of the four red giants located in the Hyades.}
   {We analysed time-series data from both ground and space to perform the asteroseismic analysis. By combining high signal-to-noise (S/N) 
   radial-velocity data from the ground-based SONG network with continuous space-based data from the revised {\it Kepler} mission K2, 
   we derive and characterize 27 individual oscillation modes for $\epsilon$\,Tau, along with global oscillation parameters such as the large frequency
   separation $\Delta\nu$ and the ratio between the amplitude of the oscillations measured in radial velocity and intensity as a function of frequency. The latter has
   been measured previously for only two stars, the Sun and Procyon. Combining the seismic analysis with interferometric and spectroscopic 
   measurements, we derive physical parameters for $\epsilon$\,Tau, and discuss its evolutionary status.}
   {Along with other physical parameters, we derive an asteroseismic mass for $\epsilon$\,Tau of $M=2.458\pm0.073\, \rm M_{\odot}$, which is slightly lower than previous estimates, and which leads to a revised minimum mass of the planetary companion. Noting that the SONG and K2 data are non-simultaneous, we estimate the amplitude ratio between intensity and radial velocity to be $42.2\pm2.3$\,ppm\,m$^{-1}$\,s, which is higher than expected from scaling relations.}
   {}

   \titlerunning{Asteroseismology of $\epsilon$\,Tau}
   \authorrunning{T. Arentoft, F. Grundahl, T. R. White et al.}
   \maketitle
%

\section{Introduction}
Stellar open clusters are testbeds for stellar astrophysics because the common 
distance, chemical composition, formation history and age of the stars in a 
cluster 
limits the number of free parameters when fitting models to multiple cluster members.
Furthermore, clusters 
often include objects that provide even more detailed information, such as 
stars in eclipsing binary systems, stars with exoplanets, and oscillating 
stars.  Nearby clusters offer even better prospects, as they can be studied 
using multiple complementary techniques, including interferometry and time-resolved 
spectroscopic observations. 

As the nearest open cluster, the Hyades is very well-studied, and even a 
casual inspection of the literature shows it to be an important laboratory for 
studying stellar evolution and stellar properties in great detail.
With an age around 650\,Myr \citep{lebreton2001} the cluster stars span a large 
range in mass. The highest masses are represented by the four brightest giants 
($\gamma$, $\epsilon$, $\theta^1$, and $\delta^1$\,Tau), which are all thought to be
in the core-helium-burning stage \citep{debruijne2001} based on their location
in the cluster colour--magnitude diagram. If this is indeed the case, $\epsilon$\,Tau, which is the 
subject of this paper, would belong to the secondary clump given its mass of around 2.5\,M$_{\odot}$ 
\citep{Girardi99,Montalban13}. $\epsilon$\,Tau is furthermore a known exoplanet host, with a massive 
planet ($m_{2}\,\rm{sin}\,i = 7.6\pm 0.2\, \rm M_{\rm J}$) in an 595-d orbit \citep{Sato2007}.

The revised {\it Kepler} mission, K2 \citep{Howell14}, has uncovered 
several planetary systems in the cluster \citep{Mann2018, Ciardi2018, Livingston2018}.  
The high-precision photometry from K2 has also allowed the detection of solar-like oscillations in 
two main--sequence stars \citep{Lund2016} and in the four bright giants (White et al., in prep.).  
Prior to this, \citet{Ando2010} detected oscillations in $\epsilon$\,Tau based on a few nights of data, with an oscillation 
signal which is in good agreement with our analysis below. Furthermore, \cite{Beck2015} reported the detection of oscillations 
in $\theta^1$\,Tau. 

We have used the two nodes of the
Stellar Observations Network Group (SONG) to obtain high-precision radial 
velocities from the sites in Tenerife \citep{grundahl2017} and the 
Delingha observing station in China \citep{Deng13} for $\epsilon$\,Tau. The goal was to provide a dataset of 
radial velocities of the same duration as the K2 photometric data, such that a
combined analysis could be carried out. Space-based photometry offers
uninterrupted observations over long time spans and thus 
provide a good window function for the frequency analysis, whereas radial-velocity (RV)
observations have a much higher sensitivity to the oscillations. This is because the background from 
stellar granulation is much lower in RV than in photometry \citep{Bedding06,Garcia13}.
Since the RV observations are ground-based, the window function is not as
good as from K2, which complicates the detection of true oscillation modes versus
aliases. 
Thus, in the ideal case the combination of space- and ground-based observations
will allow a correct identification of oscillation modes and provide a high S/N
ratio. 

With the seismic data presented here, we provide an updated estimate
of the evolutionary state and mass for $\epsilon$\,Tau and a precise value 
for its surface gravity. From our high-resolution spectra we derived the
effective temperature and abundances. 
We also measured the ratio between the photometric and RV amplitudes as a function of frequency, 
an important measurement that can be used to test models of the stellar atmosphere \citep{Houdek10}.
To our knowledge this represents only the second star with solar-like oscillations (apart from the Sun itself)
with such a measurement; the other being Procyon \citep{Huber11}.   

\section{Observations and data reduction}

As part of this project, $\epsilon$\,Tau was, as mentioned, observed with the Hertzsprung SONG telescope
in Tenerife \citep{grundahl2017} and the Chinese SONG telescope at the Delingha Observing 
Station \citep{Deng13}.
We also include data from the revised {\it Kepler} mission K2 \citep{Howell14}, 
taken during campaign 13 in which 
$\epsilon$\,Tau was observed. 

\subsection{Tenerife data}

The 1\,m Hertzsprung SONG telescope on Tenerife observes in an fully automated mode \citep{MFA16}, 
and all the $\epsilon$\,Tau data were collected in this way. The frequency of maximum
oscillation power, $\nu_{\rm max}$, is around 60\,$\mu$Hz, 
or $\sim$4.5\,hrs \citep[see preliminary results 
presented by][]{stello2017}, and we therefore allowed
other observing programmes to be executed as short (typically 1 hour) interruptions of the
$\epsilon$\,Tau time-series observations two or three times per night.  We used an iodine cell 
for precise wavelength calibration, providing a single-point precision of 2--3\,m\,s$^{-1}$. The 
observation of the needed stellar template, spectral extraction and velocity calculations 
followed closely the method employed in \citet{grundahl2017}. 
For constructing the stellar template we obtained nine spectra with the highest spectral 
resolution (110 000), resulting in a S/N ratio above 300 at wavelengths
longer than 5000{\AA} (the spectrograph has a spectral range of 440\,nm--690\,nm). 
This combined spectrum was used for the abundance analysis presented
in Sec.~\ref{spec_analysis}. In total, 5766 spectra were obtained for 
$\epsilon$\,Tau (see Table~\ref{table:1}). The first 941 spectra were obtained in November 2015 \citep{stello2017}, 
while the majority of the spectra (4825) were obtained in the period of October 2016 to 
January 2017. Only the latter data were used in our asteroseismic analysis. 
We discarded a few outliers and ended up with a time series from Tenerife consisting of 4811 
RV measurements for $\epsilon$\,Tau. 
 
\subsection{Delingha data}

The Delingha site is the second SONG node. It has a 1\,m diameter telescope and
as for the Tenerife node, the
main instrument is a high-resolution \'echelle spectrograph (designed with the same throughput and
resolution) located at the coud\'e focus. The building is insulated and temperature-controlled, since
the temperature variation at the site can be somewhat larger than at Tenerife. 
In general, the site has slightly poorer weather conditions than the Tenerife site. 
The best observing conditions occur during the September--April season. 

The spectrograph is very similar to the one in Tenerife; the main difference is that the spectral
range is slightly smaller (440\,nm--680\,nm), with 50 spectral orders. The iodine cell uses
counter-rotated wedged end windows to avoid fringing in the stellar spectra. The calibration, 
spectral extraction and velocity calculation are done by the same software as for the 
Tenerife data. 

Since the operation of this telescope is not yet automated, all observations were carried out 
with an observer present. The RV precision obtained is 
close to 4\,m\,s$^{-1}$, somewhat worse than for Tenerife. This is probably due to non-automatic guiding 
and a lower signal due to the seeing. We do, however, note that during individual nights with the 
best observing conditions, a single-point precision better than 3\,m\,s$^{-1}$ has been reached.
We collected in total 849 RV measurements from China; however, the data from some of the 
nights were of low quality and have been omitted in the analysis. We retained 590 RV measurements for 
use in the asteroseismic analysis. 

\subsection{The combined SONG data}

   \begin{figure}
   \centering
   \includegraphics[width=9cm]{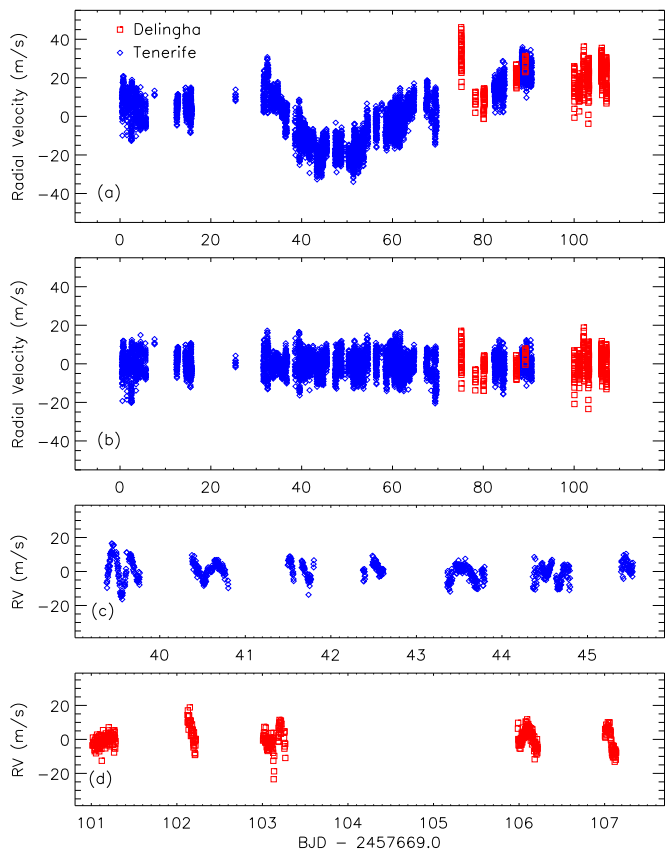}
     \caption{SONG data from Tenerife and Delingha, (a) the combined, unfiltered time-series
              (b) the combined, filtered time-series. Panels (c) and (d) show close-ups of individual
              nights of RV observations from Tenerife (c) and from Delingha (d). 
             }
         \label{Fig.SONGTS}
   \end{figure}

The data from Tenerife and Delingha were combined by shifting each series to a common RV zero-point. The
combined time series can be seen in Fig.\,~\ref{Fig.SONGTS}a. There is a slow 
variation with an amplitude of about 20\,m\,s$^{-1}$ clearly visible in the data; similar drifts are visible in 
other SONG time-series and are most likely instrumental, although part of the variations may also be
caused by rotational modulation or stellar activity, as they resemble variations seen in the RV data of $\theta^1$\,Tau by \citet{Beck2015}. The length of our time series is less than 20\%
of the orbital period of the exoplanet, so we expect only very-low-frequency modulation caused by the planet. We therefore filtered the data by 
subtracting a number of dominating, low-frequency sinusoidal signals, resulting in the combined time-series 
seen in Fig.~\ref{Fig.SONGTS}b. The two lowest panels in Fig.\,~\ref{Fig.SONGTS} are close-up views 
from Tenerife and Delingha, respectively. Oscillations with a period of roughly 0.2\,d are clearly 
visible in the data from both telescopes.

\subsection{K2 data}
The K2 mission
observed $\epsilon$\,Tau during Campaign 13 (2017 March 8 to May 27) under Guest Observer Programme 13047 (P.I. D. Huber), with an observing cadence of one measurement per 30 minutes. Due to its brightness, $\epsilon$\,Tau saturates the \emph{Kepler} detector, with excess flux bleeding along CCD columns. 
Due to limitations of on-board data storage and telemetry from \emph{Kepler}, it is not practical to record all the flux
from such a bright star because of the large number of pixels this would require. Instead, $\epsilon$\,Tau was observed with a circular mask with a radius of 20 pixels, with the time-series constructed from a weighted sum of the unsaturated pixels in the halo of scattered light surrounding the star. This method, referred to as `halo' photometry, removes trends in the time-series that are due to the drift of the telescope, and has been successfully demonstrated with K2 observations of the bright B-stars in the Pleiades \citep{White2017}, as well as the red giant Aldebaran \citep{Farr2018}.

\begin{figure}
   \centering
   \includegraphics[width=9cm]{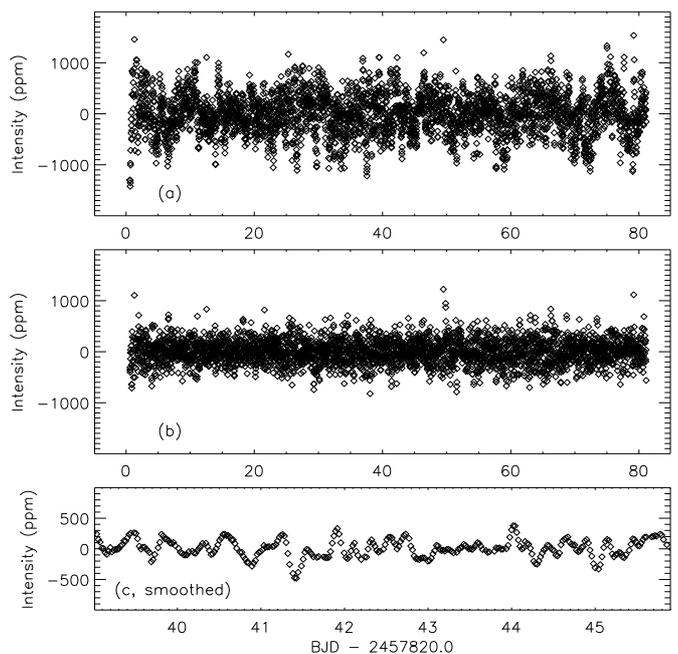}
     \caption{K2 data for $\epsilon$\,Tau. (a) the unfiltered time-series, (b) a version of the time series 
              where dominating low-frequency variations up to 42\,$\mu$Hz have been subtracted (see text), (c) 
              a zoomed view of the filtered data for the same length in time as in Fig\,\ref{Fig.SONGTS}c (we note the different zero points
              in time between the SONG and the K2 data). The data shown in panel c have been smoothed to enhance the oscillation signal. 
             }
         \label{Fig.K2TS}
   \end{figure}

The K2 time-series, which can be seen in Fig.\,\ref{Fig.K2TS}a, has an average precision per 
point of 218\,ppm. The dominating low-frequency variations were again subtracted, providing us with the filtered
time-series seen in panels (b) and (c) of Fig.\,\ref{Fig.K2TS}.
In the filtered time-series, the average noise per point is 160\,ppm. 

\begin{table}
\caption{Observational data from Tenerife and Delingha, and from K2.}
\label{table:1}
\centering
\begin{tabular}{l c c c }
\hline\hline
Parameter & Tenerife & Delingha & K2 \\
\hline
   First data             & 2015-11-18 & 2016-12-11 & 2017-03-08  \\
   Last data              & 2017-01-16 & 2017-01-25 & 2017-05-27  \\
   N$_{\rm exposure}$     & 5766       & 849        &  3390   \\
   Exposure time (s)      & 180        & 240        & $\sim$1800    \\
   Spec. resolution       & 77\,000    & 77\,000    & --    \\ 
   RV error (m\,s$^{-1}$) & 2.8        & 4.0        & --    \\
\hline                              
\end{tabular}
\end{table}

\section{Data analysis}
\subsection{Power spectra and global oscillation parameters}
\label{sec:power-spectrum}

\begin{figure}
   \centering
   \includegraphics[width=9cm]{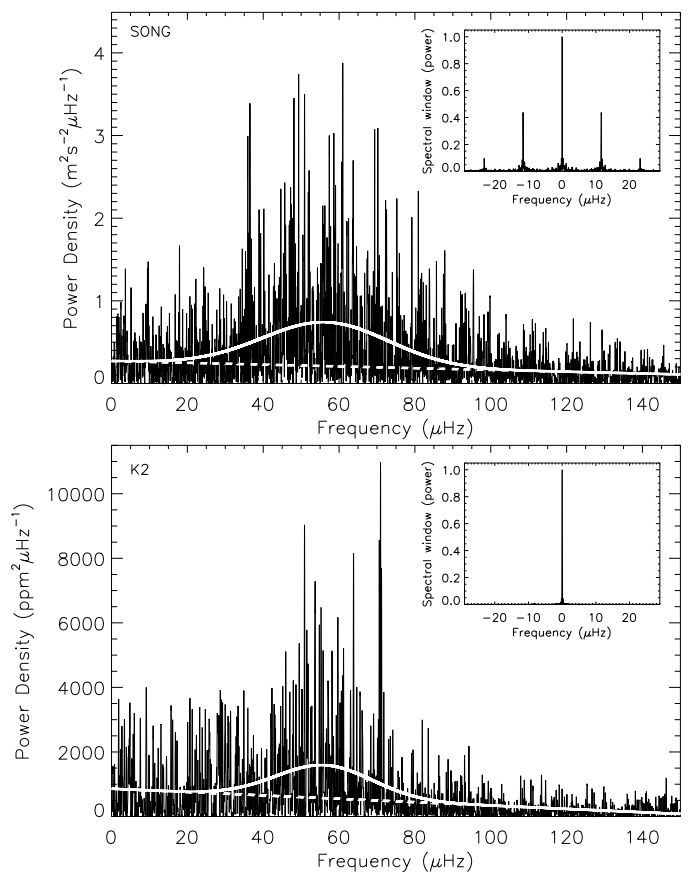}
     \caption{Power density spectra and spectral windows (in power) for the SONG and K2 data, respectively. The white curves are fits to the data to determine $\nu_{\rm max}$, see text for a discussion. 
             }
         \label{Fig.PS}
   \end{figure}

The power spectra of the SONG and K2 time-series data were calculated using unweighted sine wave fitting to the data \citep{Frandsen95}. We did try to use statistical weights for the SONG data, based on either the overall noise level in the data from each of the two sites, the flux levels of the individual measurements, or the local scatter in the time-series (obtained by running a boxcar through a version of the time series where all signals, including the oscillations, had been removed). We were only able to obtain a marginal improvement in the white-noise level as a result of the statistical weights, at the cost of degrading the spectral window by downweighting the data from China. We ascribe the ineffectiveness of using weights to the fact that the SONG data are dominated by the data from Tenerife, which are very homogeneous. Statistical weights were therefore not used in our analysis. The power density spectra and the corresponding spectral windows are shown in Fig.\,\ref{Fig.PS}. 

Although the SONG data were obtained 
from two sites, the final time-series still contains gaps, which result in $1\,\rm{d}^{-1}$ sidelobes of about 40\% in power
(65\% in amplitude), which complicates the frequency analysis. The spectral window
for the K2 data is excellent; there is, however, a significant rise in power at low frequencies due to 
stellar granulation. 
The oscillation envelope is clearly visible in both power spectra, with a frequency of maximum power ($\nu_{\rm max}$) just below 60\,$\mu$Hz. The white-noise level in amplitude at high frequencies (above 500\,$\mu$Hz) is 7.0$\,\rm{cm}\,{\rm s}^{-1}$ for the SONG data, translating to an average noise of 2.9$\,\rm{m}\,{\rm s}^{-1}$ per data point. This is in good agreement with the noise estimates in Table\,\ref{table:1} and given that the combined dataset is dominated by the data from Tenerife. The white-noise level in the K2 spectrum is 4.88$\,{\rm ppm}$, which translates into the noise per data-point of 160$\,{\rm ppm}$ in the filtered time series, as mentioned above. 

Following the approach of \citet{Mosser2009}, which was also applied to SONG data by Stello et al. (2017), the frequency of maximum power ($\nu_{\rm max}$) was determined from the SONG $\epsilon$\,Tau data by applying a Gaussian fit combined with a linear trend to the filtered data, in order to take background signals into account. This is shown in the upper panel of Fig.\,\ref{Fig.PS}; the full white line is the combined linear and Gaussian fit, while the dashed line is the linear fit to the background. From these fits, the frequency of maximum power was found to be 56.4\,$\mu$Hz. 
The oscillations are stochastic, which means that the distribution of oscillation power (i.e. which modes have highest amplitude) will differ from one instant to the next, providing slightly different values for $\nu_{\rm max}$ depending on when the star is observed. We used the actual SONG time-series to quantify this effect, and hence to estimate the uncertainty on $\nu_{\rm max}$. 
We did this in the following way: we created 10 versions of the time series in each of which we had subtracted one of the dominant oscillation signals using CLEAN (the details of this method are described in Sec.~\ref{sec:clean} below). In order to avoid a bias, the 10 subtracted signals were evenly distributed around $\nu_{\rm max}$, in a range between 35.93 and 87.79\,$\mu$Hz. We then calculated the power spectra based on these 10 modified time series, and determined $\nu_{\rm max}$ for each one of them.
In this way we found 10 slightly different $\nu_{\rm max}$ values, which turned out to have a mean value of 56.4\,$\mu$Hz, and a standard deviation of 1.05\,$\mu$Hz which we round up to 1.1\,$\mu$Hz and adopt as our uncertainty estimate on $\nu_{\rm max}$, as reported in Table\,\ref{table:2}. To test if this uncertainty is realistic, we used a 4-year {\it Kepler} time-series of KIC9716522, which is very similar to $\epsilon$\,Tau, see \citet{Arentoft17}, but with a clearer oscillation signal, so that an 80-d segment of the KIC9716522 data resembles more our $\epsilon$\,Tau RV data than the $\epsilon$\,Tau K2 data where the oscillation signal is less pronounced. We split the time series into 18 80-d segments, filtered each of the segments for low-frequency signals, as we have done for the $\epsilon$\,Tau data, and determined $\nu_{\rm max}$ in the same way as for $\epsilon$\,Tau. In this way we found 18 values of $\nu_{\rm max}$ for KIC9716522, which turned out to have a standard deviation of 0.9\,$\mu$Hz, in good agreement with our estimated uncertainty of 1.1\,$\mu$Hz for $\epsilon$\,Tau. Although the spectral windows of these 18 space-based data segments are cleaner than
the SONG spectral window, this test supports our estimated uncertainty for the $\epsilon$\,Tau 
$\nu_{\rm max}$ value of 1.1\,$\mu$Hz.

Finally, we also performed the Gaussian fit to the K2 data and show the results in the lower panel of Fig.\,\ref{Fig.PS}; the value of $\nu_{\rm max}$ determined in this way is 56.1$\pm2.4$\,$\mu$Hz, in good agreement with the value of 56.4$\pm$1.1\,$\mu$Hz found from the SONG data. In this case the uncertainty 
was found following the approach of \citet{Arentoft17}; the K2 time series was split in two, $\nu_{\rm max}$ was found from each of the two halves series, and the uncertainty was taken as the 
difference between these two values divided by $\sqrt{2}$. Given the higher S/N
in the spectroscopic data, we retain the value from SONG as our final result for $\nu_{\rm max}$.

\begin{table}
\caption{Global asteroseismic parameters for $\epsilon$\,Tau.}
\label{table:2}
\centering
\begin{tabular}{l c c}
\hline\hline
Parameter & SONG & K2  \\ 
\hline                        
   $\nu_{\rm max}$      & 56.4$\pm$1.1\,$\mu$Hz   & 56.1$\pm$2.4\,$\mu$Hz  \\  
   $\Delta\nu$          & 5.00$\pm$0.01\,$\mu$Hz  & 5.00$\pm$0.1\,$\mu$Hz \\
   $\delta\nu_{\rm 02}$ & 0.76$\pm$0.05\,$\mu$Hz    & \\
   $\epsilon$           & 1.19$\pm$0.06             & \\
\hline                                 
\end{tabular}
\end{table}

   \begin{figure}
   \centering
   \includegraphics[width=9cm]{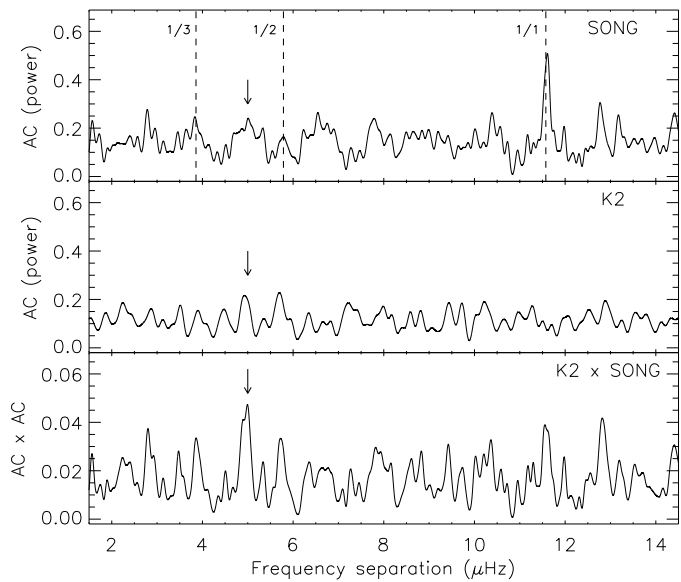}
     \caption{Autocorrelations of the power spectra, for the SONG spectrum in the upper panel, and K2 in the middle panel. The bottom panel shows the two autocorrelations multiplied together. The vertical dashed lines in the upper panel indicate the signals originating from the 1\,${\rm d}^{-1}$ aliases in the spectral window, at 11.574\,$\mu$Hz and integer fractions thereof. The arrows indicate a value of 5.0\,$\mu$Hz, which we identify as $\Delta\nu$ for $\epsilon$\,Tau using various methods (see text).  
             }
         \label{Fig.AC}
   \end{figure}

   \begin{figure}
   \centering
   \includegraphics[width=9cm]{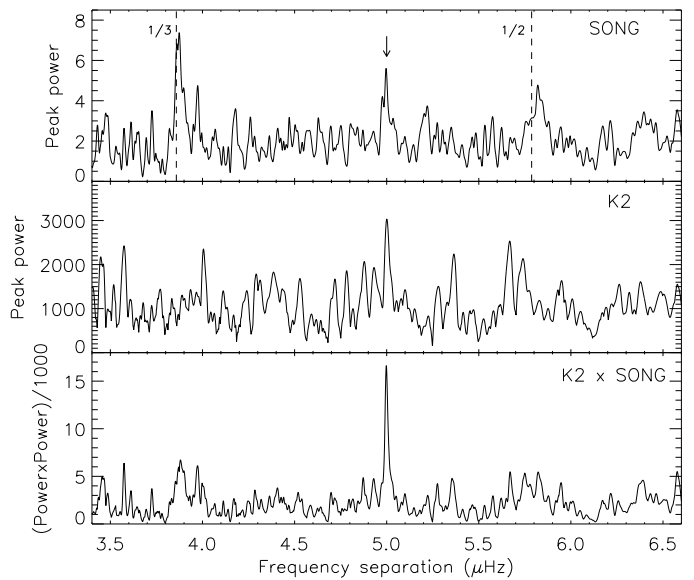}
     \caption{Results of using a modified version of an analysis method developed for solar-like oscillations observed with {\it Kepler}, as described in the text. The upper panel shows the results based on the SONG power spectrum, and the middle panel shows the results for K2. As in Fig\,\ref{Fig.AC}, the dashed lines in the upper panel indicate signals originating from the 1\,${\rm d}^{-1}$ aliases in the spectral window, and the arrow indicates the $\Delta\nu$-value of 5.0\,$\mu$Hz. In the bottom panel, the SONG and K2 results from the two upper panels are multiplied, resulting in a very clear peak near 5.0\,$\mu$Hz.
             }
         \label{Fig.PowSum}
   \end{figure}

We then proceeded to determine the large frequency separation $\Delta\nu$, which is the frequency separation between oscillation modes of consecutive radial order ($n$) with the same angular degree ($\ell$), assuming that the oscillations follow the asymptotic relation (see Sec.~\ref{sec:clean}). Because of the expected regularity of the frequency spectrum, we first looked at the autocorrelation of each of the two power spectra, based on the SONG- and K2-data. The results are shown in Fig.\,\ref{Fig.AC}. The autocorrelation of the SONG power spectrum is, not surprisingly, dominated by a peak from the 1\,${\rm d}^{-1}$ aliases originating from the spectral window. There is a minor peak at 5.0\,$\mu$Hz in both autocorrelations, marked by an arrow in Fig.\,\ref{Fig.AC}; this peak becomes more prominent when we multiply the two autocorrelations, as seen in the lower panel. This is a signature of the large frequency separation for $\epsilon$\,Tau, but we only have a marginal detection of $\Delta\nu$ based on the autocorrelation method alone.

Instead we applied a method described in \citet{JCD08}, and in \citet{Arentoft17} for the stars in NGC\,6811, with a slight modification. For each trial $\Delta\nu$ value in a range where we expect to find $\Delta\nu$ for $\epsilon$\,Tau, we cut the region of the SONG and K2 power spectra (respectively) containing the oscillations in bins of $\Delta\nu$, added the bins and found the highest peak in the summed spectrum. In this way, modes of $\ell$=0 will add up in the summed spectrum and create a strong peak when the correct value for $\Delta\nu$ is used. We have previously cut the spectrum in bins of $\Delta\nu$/2, in order to make modes of $\ell$=0,1 to add up. 
However in evolved stars like $\epsilon$\,Tau, the $\ell$=1 modes are expected to be mixed \citep{Dziembowski2001,JCD2004,Dupret2009} and their frequencies will therefore deviate from the values expected from the asymptotic relation for pure p-modes (Eq. \ref{eq.asymtotic}). We show the results of this analysis in Fig.\,\ref{Fig.PowSum}. The figure plots the summed value for the highest peak as a function of the trial $\Delta\nu$-value. In the SONG data, we again see peaks at fractions of the daily aliases, but also a relatively strong peak near 5.0\,$\mu$Hz. The same peak is present in the K2 data, and when multiplying the two (bottom panel), we obtain a very clear peak just slightly below 5.0\,$\mu$Hz which we interpret as the large frequency separation. 
In the following, we identify a number of modes separated almost exactly by 5.0\,$\mu$Hz. We assume these to be equidistant $\ell$=0 modes and use these modes below to refine the $\Delta\nu$-value for $\epsilon$\,Tau, based on the SONG data alone. The uncertainty value quoted in Table\,2 for $\Delta\nu$ of 0.01\,$\mu$Hz for SONG originates from this analysis, which is presented in the next section. The uncertainty value for K2 was again found by splitting the K2 time series in two, repeating the analysis on the two half series, and taking the uncertainty as the difference between the two values, divided by $\sqrt{2}$. 

\subsection{Individual frequencies}
\label{sec:clean}

The individual oscillation frequencies were found from the SONG data. We then used the K2 data, where the oscillation modes have lower S/N but a much cleaner spectral window, to distinguish between true oscillation modes and daily aliases. $\epsilon$\,Tau is an evolved star, so we assume that the modes with angular degree $\ell$=0 and $\ell$=2 will largely follow the asymptotic relation \citep{vandakurov1967,tassoul1980,gough1986};
\begin{small}
\begin{equation}
\nu_{n,\ell}\approx\Delta\nu(n+\frac{1}{2}\ell+\epsilon)-\ell(\ell+1)D_0,
\label{eq.asymtotic}
\end{equation}
\end{small}
\noindent
while there will be multiplets of mixed $\ell$=1 modes that do not follow this relation. The frequency analysis was an iterative procedure where we first ran a simple CLEAN \citep{Frandsen95} on the SONG data, where the dominating modes are subtracted one by one from the time series, with the power spectrum being recomputed in each step, and with the criterion that modes were included in the 
frequency list if their amplitudes were above 3.0 times the mean level of the amplitude spectrum between 130 and 150\,$\mu$Hz. All our detected modes have frequencies below 100\,$\mu$Hz. The average noise level in this part of the spectrum is more than three times higher than if we had used the noise level at even higher frequencies (e.g. above 500\,$\mu$Hz), which is likely due to spectral leakage through the window function of undetected modes in the frequency range where we detect the modes. However, finding the noise estimate close to the oscillation frequencies gives a more realistic estimate of the noise level in the region where we detect the modes, even if part of the noise is due to undetected modes, and results in a conservative frequency analysis and conservative estimates for the final S/N values of the detected modes. 

   \begin{figure}
   \centering
   \includegraphics[width=9cm]{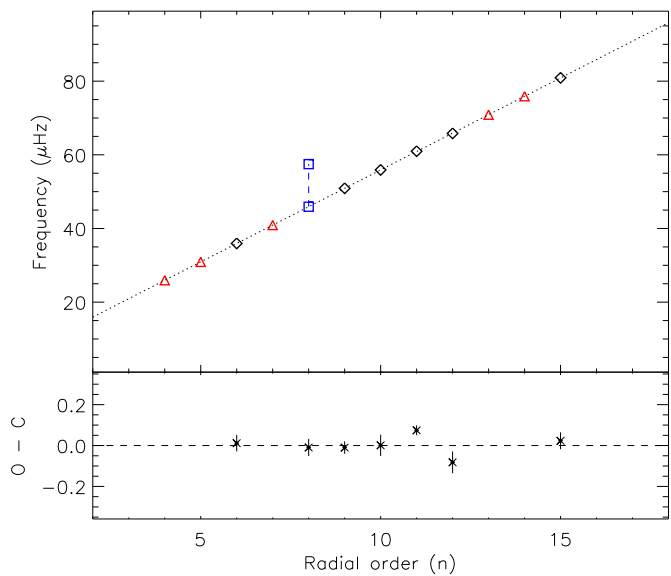}
     \caption{Upper panel: Frequencies as a function of radial order ($n$) for the first stage of our frequency analysis. The dotted line has a slope of 5.00\,$\mu$Hz, in agreement with $\Delta\nu$ found in Fig.\,\ref{Fig.PowSum}. The black diamonds are detected modes that fit into a regular structure of expected $\ell$=0 modes, red triangles mark the positions of expected $\ell$=0 modes that were undetected in our first run of CLEAN, and the two blue squares represent the detected $\ell$=0, $n$=8 mode, for which the first run of CLEAN picked up the +1\,d$^{-1}$ alias (see text). The lower panel shows the differences between the detected $\ell$=0 frequencies and those predicted by the asymptotic relation (Eq.\ref{eq.asymtotic}), also in $\mu$Hz.  
             }
         \label{Fig.l0fit1}
   \end{figure}

\begin{figure}
   \centering
   \includegraphics[width=9cm]{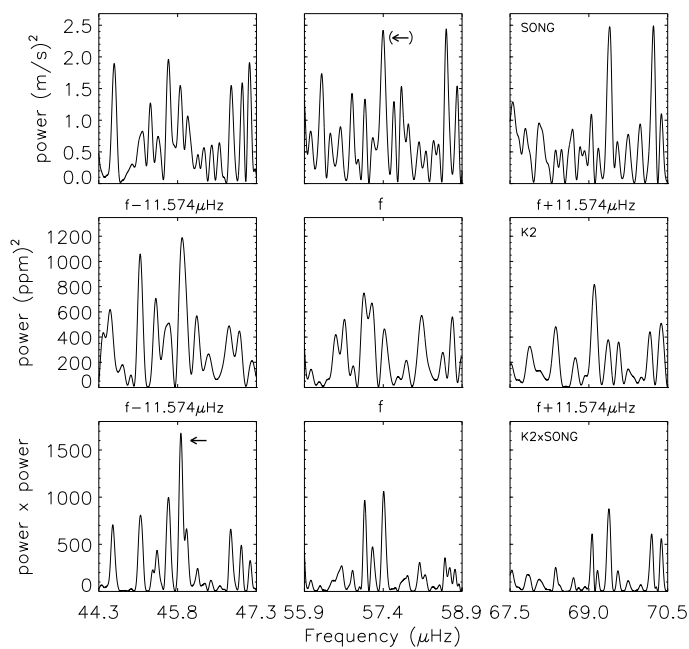}
     \caption{As shown in Fig.\,\ref{Fig.l0fit1}, one of the modes detected in the original run of CLEAN on the SONG data was likely an 1\,d$^{-1}$ alias of a true $\ell$=0 mode. The three upper panels show SONG data near the originally detected frequency (middle column of panels), and minus (left) and plus (right) 1\,d$^{-1}$ corresponding to 11.574\,$\mu$Hz. The middle panels show the same for the K2 data, and the lower panels for the product of the two. The K2 power spectrum and the product of the SONG and K2 spectra support the interpretation that the originally detected mode was in fact a 1\,d$^{-1}$ alias of the actual mode, which is indicated by an arrow in the lower-left panel. 
             }
         \label{Fig.fz}
   \end{figure}

At this point of our analysis we could potentially have detected peaks that are daily aliases of true oscillation modes, which would mean that we obtain wrong frequency values. As a result, the CLEAN procedure would become sub-optimal when we subtract these wrong frequencies during the analysis. We therefore looked for a regular pattern among the 20 frequencies detected in this first run of CLEAN, to identify a series of regularly spaced $\ell$=0 modes. This turned out to be possible. In Fig.\,\ref{Fig.l0fit1} we show a number of modes which are all separated by 5.00\,$\mu$Hz, in good agreement with the value of the large frequency separation, $\Delta\nu$, found above, and to which we could assign a radial order ($n$). In Fig.\,\ref{Fig.l0fit1} the black diamonds indicate detected modes that fit this regular structure, while the red triangles mark the positions of expected $\ell$=0 modes that were undetected in the first run of CLEAN. The blue squares indicate a mode for which CLEAN most likely picked up a 1\,d$^{-1}$ alias. The original frequency at 57.415\,$\mu$Hz lies almost exactly 11.574\,$\mu$Hz above the predicted $\ell$=0 mode with $n$=8. To check this interpretation, we looked at the power spectra based on SONG, K2 and SONG multiplied by K2 near the detected frequency, and its $\pm$\,1d$^{-1}$ aliases. This is shown in Fig.\,\ref{Fig.fz}. The SONG data shown in the upper panels favour the originally detected frequency in the middle panel, while the K2 data and the product of the SONG and K2 data support that the true mode is the one near 45.8\,$\mu$Hz, which also fits into the regular $\ell$=0 structure shown in Fig.\,\ref{Fig.l0fit1}. The 1d$^{-1}$ alias is probably dominating in the SONG spectrum because its amplitude is influenced by nearby modes, or by the spectral window of adjacent modes.

\begin{table}
\caption{List of the 27 mode frequencies (with uncertainties) in $\epsilon$\,Tau, their amplitude, S/N value in amplitude (not power), and mode-identification ($\ell,n$) for the $\ell$=0,2 modes. The last column lists the abscissa coordinate used in the {\'e}chelle diagram in Fig.\,\ref{Fig.echelle}.}
\label{table:3}
\centering                        
\begin{tabular}{c c c c c c} 
\hline\hline                 
f\,($\mu$Hz) & $\sigma(f)$      & a (m/s) &     S/N (a) &   Mode ID &           {\'E}chelle abs. \\ 
\hline                        
       17.92 &      0.05 &        1.12 &        6.12 &                &        3.23 \\
       20.68 &      0.06 &        0.93 &        5.08 &                &       0.99 \\
       31.55 &      0.06 &        0.94 &        5.13 &                &        1.87 \\
       35.93 &      0.04 &        1.37 &        7.52 & $\ell$=0,$n$=6 &        1.26 \\
       38.86 &      0.06 &        0.78 &        4.26 &                &        4.18 \\
       40.17 &      0.05 &        1.20 &        6.59 & $\ell$=2,$n$=6 &       0.50 \\
       41.52 &      0.06 &        0.72 &        3.93 &                &        1.85 \\
       44.63 &      0.06 &        0.89 &        4.89 &                &        4.96 \\
       45.90 &      0.04 &        1.32 &        7.25 & $\ell$=0,$n$=8 &        1.23 \\
       47.23 &      0.05 &        1.17 &        6.40 &                &        2.56 \\
       48.10 &      0.04 &        1.26 &        6.88 &                &        3.43 \\
       50.90 &      0.03 &        1.68 &        9.20 & $\ell$=0,$n$=9 &        1.23 \\
       51.98 &      0.06 &        0.78 &        4.27 &                &        2.31 \\
       53.26 &      0.06 &        0.87 &        4.79 &                &        3.60 \\
       55.73 &      0.05 &        1.18 &        6.48 &                &        1.07 \\
       55.90 &      0.05 &        1.01 &        5.56 & $\ell$=0,$n$=10  &        1.24 \\
       57.39 &      0.05 &        1.07 &        5.84 &                  &        2.73 \\
       60.97 &      0.02 &        1.87 &        10.26 & $\ell$=0,$n$=11 &        1.32 \\
       63.70 &      0.04 &        1.52 &        8.34 &                  &        4.04 \\
       65.81 &      0.05 &        0.98 &        5.35 & $\ell$=0,$n$=12  &        1.16 \\
       70.85 &      0.07 &        0.46 &        2.52 & $\ell$=0,$n$=13  &        1.20 \\
       75.87 &      0.06 &        0.79 &        4.32 & $\ell$=0,$n$=14  &        1.22 \\
       79.20 &      0.06 &        0.84 &        4.60 &                  &        4.56 \\
       80.90 &      0.04 &        1.34 &        7.34 & $\ell$=0,$n$=15  &        1.26 \\
       83.00 &      0.07 &        0.59 &        3.25 &                  &        3.36 \\
       83.79 &      0.05 &        1.09 &        5.98 &                  &        4.15 \\
       96.34 &      0.06 &        0.73 &        3.98 &                  &        1.71 \\
\hline                                   
\end{tabular}
\end{table}

\begin{figure}
   \centering
   \includegraphics[width=9cm]{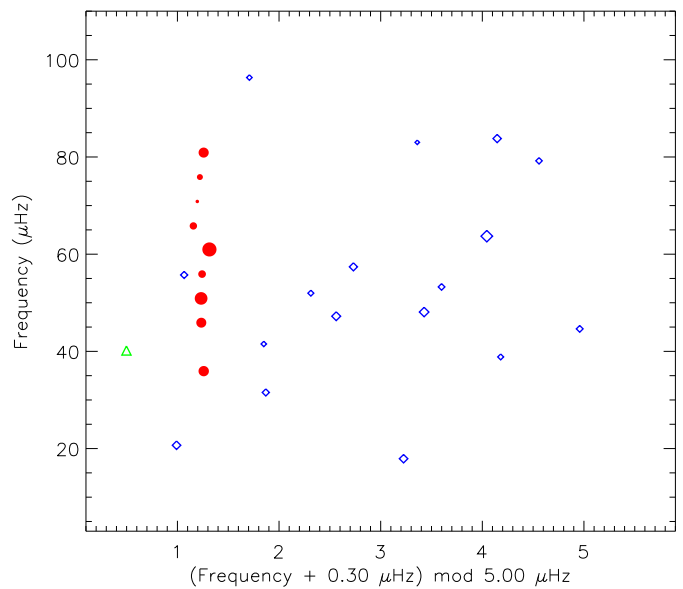}
     \caption{{\'E}chelle diagram for the 27 oscillation modes. The x-axis was shifted to place 
              $\ell$=0,2 modes close to each other in the diagram. Filled red circles are $\ell$=0   
              modes, the green triangle is $\ell$=2, and blue diamonds are for the remaining peaks, of which most are likely $\ell$=1 modes. 
             }
         \label{Fig.echelle}
   \end{figure}

\begin{figure*}
   \centering
   \includegraphics[width=18cm]{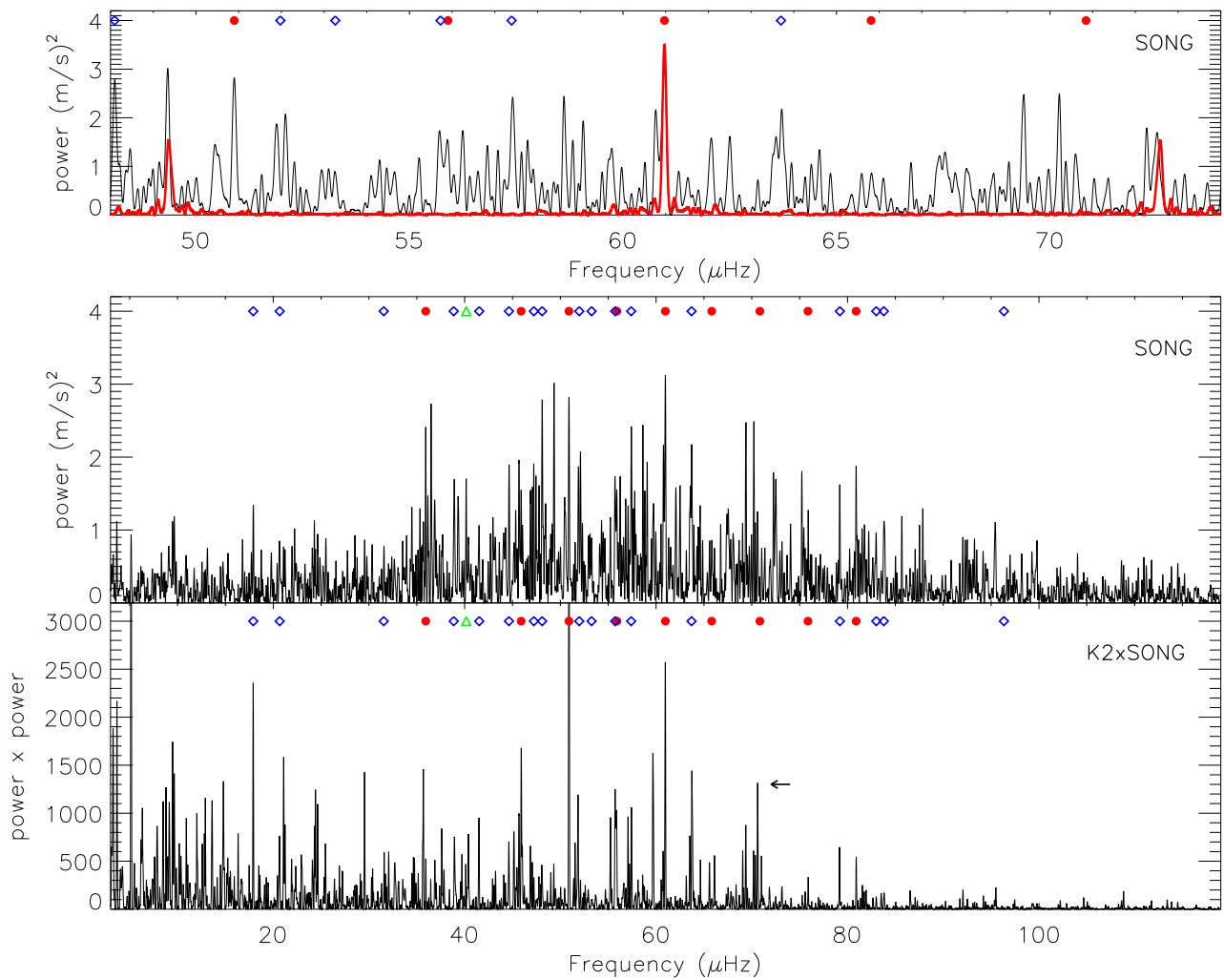}
     \caption{Top panel: A close-up of the power spectrum based on the SONG data, centred at the $\ell$=0, $n$=11 mode just below 61\,$\mu$Hz. The over-plotted, red curve shows the spectral window 
              for this central $\ell$=0 mode. The detected modes are indicated at the top of the panel,   
              with an identification as in Fig.\,\ref{Fig.echelle}.
              The $\ell$=0 mode above 70\,$\mu$Hz has a low S/N of only 2.5 in the SONG data, 
              but is more pronounced in the lower panel, where the K2 and SONG spectra are multiplied, 
              as indicated by the arrow. The mode is therefore included in our final frequency list. 
              Relatively strong but undetected peaks are sidelobes of modes outside the frequency 
              range in the close-up view. The two lower panels show the full SONG power spectrum and 
              the product of the SONG and K2 power spectra, with the detected modes indicated 
              using the same symbols as in the top panel. The three panels are aligned according 
              to the $\ell$=0 mode just below 61\,$\mu$Hz.   
             }
         \label{Fig.ShowPow}
   \end{figure*}

As a next step, we used the expected positions of the $\ell$=0 modes shown in Fig.\,\ref{Fig.l0fit1} as input to a second iteration of CLEAN on the SONG data. We included the positions of the undetected $\ell$=0 modes (the red triangles in Fig.\,\ref{Fig.l0fit1}) in order to obtain a residual time-series and power spectrum, which were unaffected by the $\ell$=0 modes and their corresponding spectral window functions. We then ran CLEAN on the residual data, in order to detect $\ell$=1,2 and possible $\ell$=3 modes. The final frequency list with all
detected modes is given in Table\,\ref{table:3}, including frequency uncertainties determined in the same way as in \citet{Arentoft17}, amplitudes, and S/N-values. We have included the mode identification based on the position in the {\'e}chelle diagram shown in Fig.\,\ref{Fig.echelle} for $\ell$=0, and for a single mode, which we identify as $\ell$=2. 
Most of the remaining modes are likely mixed $\ell$=1 modes, although some of them may be 
$\ell$=3 modes, but due to the lack of regularity among the possible $\ell$=1 modes in Fig.\,\ref{Fig.echelle}, we cannot be sure of the mode
identification for these modes. 
Some of them might also be aliases.
The detected frequencies are also marked in the power spectra shown in Fig.\,\ref{Fig.ShowPow} (see figure caption for details). 
We note that we have included the $\ell$=0, $n$=13,14 modes in our final frequency list,
although they were not detected in our first run of CLEAN. 
The $\ell$=0, $n$=14 mode has a S/N-value of 4.32 in amplitude and is therefore statistically significant, while the $\ell$=0, $n$=13 mode has a S/N-value of only 2.52 in amplitude and is not statistically significant. There is, however, evidence in the K2 data for this mode, see Fig.\,\ref{Fig.ShowPow}, and the mode has therefore been included in the final frequency list. 

\begin{figure}
   \centering
   \includegraphics[width=9cm]{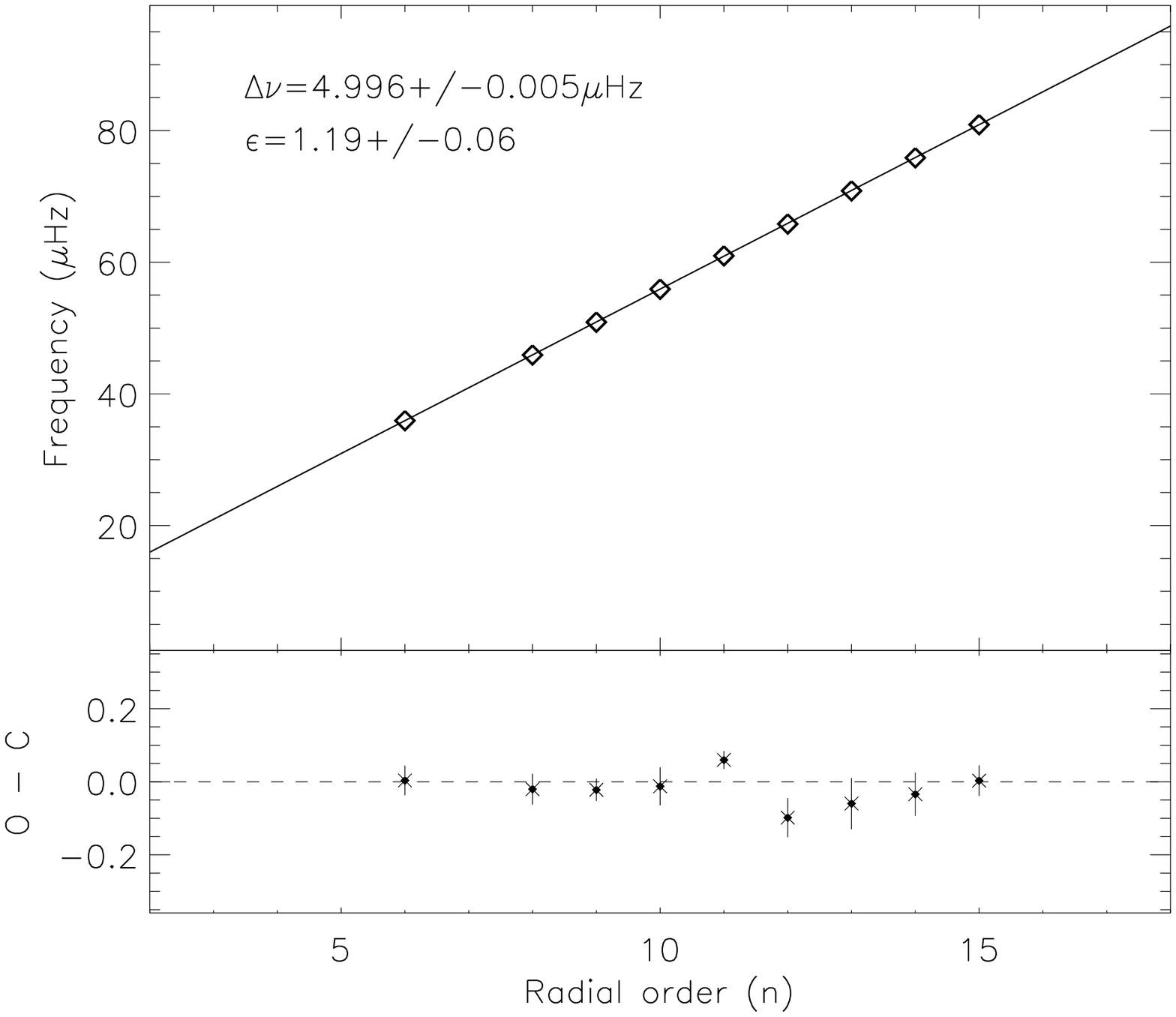}
     \caption{Frequencies of the nine $\ell$=0 modes included in Table\,\ref{table:3} as a function of their
     $n$-values. According to the asymptotic relation in Eq.\,1, the slope provides the large frequency separation $\Delta\nu$ while the intersection with the ordinate is $\Delta\nu \cdot \epsilon$, leading to the value for $\epsilon$ quoted in the figure and listed in Table\,\ref{table:2}.
             }
         \label{Fig.l0fit2}
   \end{figure}

We end up with nine regularly spaced $\ell$=0 modes, for which we plot the frequency as a function of $n$-value in Fig.\,\ref{Fig.l0fit2}. Following the asymptotic relation (Eq.\,1) we again fitted a straight line to obtain values for the large frequency separation $\Delta\nu$ and $\epsilon$ including uncertainty estimates, as shown in Fig.\,\ref{Fig.l0fit2} and as listed in Table\,2. The value for $\Delta\nu$ is in excellent agreement with the value just below 5\,$\mu$Hz found in the previous section. This is, however, not surprising, as the methods applied in Sec.~\ref{sec:power-spectrum} also searched for regularly spaced modes, and therefore also relies on the $\ell$=0 modes shown in Fig.\,\ref{Fig.l0fit2}. \\

\section{Atmospheric parameter analysis\label{spec_analysis}} 

The atmospheric parameters of $\epsilon$\,Tau were determined spectroscopically from an equivalent-width analysis. We used DAOSPEC \citep{Stetson2008a} to measure the equivalent widths using a line list published in \citet{Slumstrup2018}. The auxiliary programme {\it Abundance with SPECTRUM} \citep{Gray1994} was used to determine the atmospheric parameters, which are based on solar abundances from \citet{Grevesse1998} and ATLAS9 stellar atmosphere models \citep{Castelli2004}. We assumed local thermodynamic equilibrium (LTE) because non-LTE effects are expected to be negligible in this parameter regime \citep{Asplund2005,Mashonkina2011}. 
The atmospheric parameters were determined by invoking excitation and ionization equilibrium, and further requiring that [Fe/H] should have no systematic dependence on the strength of the line. Excitation equilibrium was reached by requiring the metallicity to show no trend with excitation potential, which is sensitive to the effective temperature, $T_{\text{eff}}$. Ionization equilibrium was reached by ensuring the average [Fe/H] agreed for absorption lines of different ionization stages, FeI and FeII. This is sensitive to the surface gravity, $\log g$, because FeII lines are more sensitive to pressure changes than FeI lines for this type of star. However, this balance is also affected by $T_{\text{eff}}$ and abundances of heavier elements, so several iterations were necessary. Lastly, the trend of [Fe/H] vs. the reduced equivalent width ($\log (EW) / \lambda$) of an absorption line is sensitive to the microturbulence, a fitting parameter introduced to describe broadening of absorption lines by turbulence on small scales, that cannot otherwise be included in 1D stellar atmosphere models. 

The analysis was carried out in two different ways, both with and without the strong asteroseismic constraint on $\log g$, calculated using the $\nu_\text{max}$ scaling relation \citep{Brown1991, Kjeldsen1995}:
\begin{align}
\log g = \log \left( \left( \frac{\nu_\text{max}}{3100\,\mu \text{Hz}} \right) \left( \frac{T_\text{eff}}{5777\,\text{K}} \right) ^{1/2} \right) + 4.44 \ .
\end{align}
The results are presented in Table~\ref{tab:spec_param} with only internal uncertainties. The asteroseismic $\log g$ is within the uncertainty of the spectroscopic $\log g$, which makes the two sets of results consistent. 
\begin{table}
\caption{Stellar atmospheric parameters determined from spectroscopy, with and without a surface gravity constraint from asteroseismology. The uncertainties are internal only, however systematic errors of the 
order of 100\,K on $T_\text{eff}$, 0.1 dex on $[$Fe/H] and 0.1--0.2 dex on $\log g$ are expected; this is
discussed in detail in \citet{Slumstrup2018}.}
\centering
\begin{tabular}{l c c}
\hline\hline
	Method               & $\log g$ free &  $\log g$ fixed \\
\hline
    $T_\text{eff}$ (K)    & 4940 $\pm 18$   & 4950 $\pm$ 22   \\
    $\log g$ (dex)        & 2.72 $\pm$ 0.07 & 2.67            \\
    $v_\text{mic}$ (km/s) & 1.30 $\pm$ 0.06 & 1.32 $\pm$ 0.06 \\
    $[$Fe/H] (dex)        & 0.16 $\pm$ 0.01 & 0.15 $\pm$ 0.02 \\
\hline    
\end{tabular}
\label{tab:spec_param}
\end{table}

With the set of atmospheric parameters determined without a $\log g$ constraint, the [Y/Mg] abundance was also determined as in \citet{Slumstrup2017} to be $0.21 \pm 0.05$. With a cluster age of 650\,Myr for the Hyades \citep{lebreton2001}, it leaves $\epsilon$\,Tau just outside one sigma agreement with the [Y/Mg] -- age relation for solar twins by \citet{Nissen2017}, their Fig.\,5, which was shown to also hold for lower-mass solar-metallicity red-clump giants ($\lesssim 1.5\, \rm M_\odot$) by \citet{Slumstrup2017}. It has been suggested in the literature that the Hyades could be as old as 800\,Myr \citep{Brandt2015} but this would slightly worsen the agreement with the [Y/Mg] -- age relation. To reach exact agreement, the cluster would instead have to be as young as 300\,Myr. However, it should be noted that \citet{Slumstrup2017} only tested the relation for lower-mass giants ($\lesssim 1.5\, \rm M_\odot$) and it is therefore not certain that the relation can be applied to giants of masses similar to $\epsilon$\,Tau.

\section{Physical parameters and evolutionary status}

The angular diameter of $\epsilon$\,Tau has been measured several times with different interferometric instruments at visible and infrared wavelengths \citep{vanBelle1999,Nordgren2001,Mozurkewich2003,Boyajian2009,vanBelle2009,Baines2018}. After correcting for limb-darkening, the measured values of the angular diameter, $\theta_\mathrm{LD}$, are not all in agreement, ranging from 2.41$\pm$0.11\,mas \citep{Nordgren2001} to 2.733$\pm$0.031\,mas \citep{Boyajian2009}. One possible source for the disagreement may be the adopted limb-darkening corrections, which are based on model atmospheres and may not be consistent between the different studies. Alternatively, it has been observed that a systematic bias towards larger angular diameter measurements exists when a star is under-resolved \citep{Casagrande2014b,White2018}. We adopt an angular diameter of $\theta_{\rm LD}$=2.493$\pm$0.019\,mas from recent measurements with the PAVO beam combiner at the CHARA Array (White et al., in prep). These observations have higher resolution than the previous measurements, allowing for the amount of limb darkening to be directly measured and lifting the degeneracy with the angular diameter. 

Combining the angular diameter with the Hipparcos parallax $\pi = 22.24\pm0.25$\,mas \citep{vanLeeuwen2007} gives a linear radius of 12.06$\pm$0.16\,$\rm R_{\odot}$. From the angular diameter and bolometric flux, $F_\mathrm{bol}\,=\,(1.27\pm0.06)\,\times\,10^{-6}\,\rm erg\,s^{-1}\,cm^{-2}$ \citep{Baines2018}, we find an effective temperature of 4976$\pm$63\,K. These values are in excellent agreement with the radius calculated from $T_{\rm eff}$, $V$-mag, parallax distance, and bolometric correction from \citet{Casagrande2014}.  This demonstrates excellent consistency between parameters and supports the spectroscopic $T_{\rm eff}$ of 4950\,K.
The Gaia DR2 parallax for $\epsilon$\,Tau is $\pi = 20.31\pm0.43$\,mas \citep{BrownGaia}, which is 
in poor agreement with the value and uncertainty reported for Hipparcos. The Gaia data product is not yet in its final version and for very bright stars, the full mission data will probably be needed to provide optimal results, and the present values are uncertain due to calibration issues \citep{GaiaDR2}. We will therefore not include an analysis based on the Gaia-DR2 parallax in this paper.
    
To determine the mass of $\epsilon$\,Tau, we used the asteroseismic scaling relations \citep[e.g.][]{Kallinger2010}.
We used the global asteroseismic parameters in Table 2 along with solar reference values from \citet{Handberg2017}.
For atmospheric parameters, we adopted the spectroscopic $T_{\rm eff}$\,=\,4950\,K and [Fe/H]\,=\,0.15, as derived in the previous section. 
To ensure exact agreement between the interferometric radius and that obtained from the asteroseismic scaling relations, we adjusted the $\Delta\nu$ correction factor $f_{\Delta\nu}$ in the asteroseismic scaling relations \citep[e.g.][]{Sharma2016,Brogaard2018} until we reached this agreement. This way we obtained an empirical value of the correction factor of $f_{\Delta\nu}=0.98993$ and we found the asteroseismic mass for $\epsilon$\,Tau to be $M=2.458\pm0.073\, \rm M_{\odot}$.  The uncertainty is based on propagating internal uncertainties on parallax, $\theta_{\rm LD}$, $T_{\rm eff}$, $\nu_{\rm max}$, and $\Delta\nu$.

The similarity in CMD position between $\epsilon$\,Tau and the other cluster giants suggests they are
all in the core-He-burning evolutionary phase. However, the empirical correction factor, $f_{\Delta\nu}$, we have found is in excellent agreement with the theoretically predicted $f_{\Delta\nu}$ by \citet{Rodrigues2017} if $\epsilon$\,Tau is assumed to be an RGB star (their Fig. 3, panels 6 and 7 counted from the top), but not if $\epsilon$\,Tau is in the core-He-burning phase. Complicating the issue, Fig. 6 (bottom panel) in \citet{Brogaard2018} reveals that the same value for $f_{\Delta\nu}$ is not found if one uses $T_{\rm eff}$ as reference to determine $f_{\Delta\nu}$ instead of $\nu_{\rm max}$, as in \citet{Rodrigues2017}, even if everything else is unchanged. This may be caused by a too low temperature scale for the models used by \citet{Rodrigues2017}, as also discussed by \citet{Brogaard2018}. Retaining the view from \citet{Brogaard2018} that it is better to use $\nu_{\rm max}$ than $T_{\rm eff}$ to estimate $f_{\Delta\nu}$, we find evidence based on our empirical correction factor pointing to the RGB as the present evolutionary phase of $\epsilon$\,Tau. If $\epsilon$\,Tau is actually a clump star, this would suggest that a slight offset is needed for the $f_{\Delta\nu}$ values predicted by \citet{Rodrigues2017}. 

\citet{Kallinger2012} found that H-shell-burning stars and core-He-burning stars could be distinguished using the central modes closest to $\nu_{\rm max}$ to determine $\Delta\nu_{\rm c}$ and $\epsilon_{\rm c}$ (see their Fig.\,4). This was also investigated theoretically by \citet{JCD2014}, who found the effect to be caused by differences in the convective envelopes. Using the radial oscillation modes in $\epsilon$\,Tau closest to $\nu_{\rm max}$ ($\ell$=0, $n$=9--11), we find $\Delta\nu_{\rm c} = 5.04\pm 0.02\,\mu$Hz and $\epsilon_{\rm c}=1.1\pm 0.2$. This places $\epsilon$\,Tau among the H-shell-burning stars in the upper panel in Fig.\,4 in \citet{Kallinger2012}. However, given the uncertainty on $\epsilon_{\rm c}$, the measurement is also consistent with $\epsilon$\,Tau being at the top of the distribution of core-He-burning stars, within 1\,$\sigma$.

We can also look at other asteroseismic indicators. Based only on a single $\ell$=2 mode, we have determined the small frequency separation $\delta_{\rm 02}$ to be 0.76\,$\mu$Hz. Using the lower panel in Fig.\,4 in 
\citet{Kallinger2012}, this places $\epsilon$\,Tau in a region of the diagram where we cannot discriminate between RGB and clump stars.
We would most likely be able to determine the evolutionary stage of $\epsilon$\,Tau if we could determine the asymptotic period spacing 
of the mixed $\ell$=1 modes, as this allows for a discrimination between core-He-burning and H-shell-burning red giants \citep{Bedding11}. For $\epsilon$\,Tau, however, the {\'e}chelle diagram shown in Fig.\,\ref{Fig.echelle} displays a sparse set of mixed 
$\ell$=1 modes, some of which may be daily aliases. Our attempts to determine the period spacing, following the methods described in \citet{Arentoft17}, were unsuccessful. 

We have also determined stellar properties using the BAyesian STellar Algorithm \citep[BASTA;][]{Victor15,Victor17}, fitting the asteroseismic parameters ($\Delta\nu$, $\nu_{\rm max}$), the interferometric temperature and spectroscopic metallicity, and available Strömgren photometry assuming $E(B-V)=0$ to determine distance. The Bayesian scheme points to the core-He-burning stage as the most likely
evolutionary stage of $\epsilon$\,Tau. The preferred model is slightly more massive than the value of $M=2.458\pm0.073\, \rm M_{\odot}$ found
from the asteroseismic scaling relations above; the model has a mass of $M=2.713^{+0.103}_{-0.182}\,\rm M_{\odot}$, however the 1$\sigma$ errorbars exactly touch each other in between the two mass estimates. In the model fits, there are also core-He-burning solutions near $2.5\,\rm M_{\odot}$, however with lower probability, and we note that also these model fits are sensitive to systematic shifts in the effective temperature, so if the effective temperature scale is shifted by, say 100\,K, the results of the model fits would be different. The radius and effective temperature of the preferred model solution ($12.46^{+0.18}_{-0.28}\, \rm R_{\odot}$ and $5004^{+55}_{-72}\, \rm K$, respectively) are slightly larger than the values found from interferometry above, which would suggest a slightly larger distance to $\epsilon$\,Tau, and hence a smaller value of the parallax, as also indicated by the preliminary Gaia parallax. We will have further constraints on the model comparison once the Gaia parallax of $\epsilon$\,Tau is final. The Bayesian scheme returns a model age of $600^{+150}_{-50}$\,Myr, consistent with the isochrone age of the Hyades. This is one of only a few cases where an asteroseismic age can be compared to other robust methods for age determination, another one being for the open cluster M67 \citep{Stello2016}.

Finally, we applied the method of \citet{Hon18}, which uses a neural network to classify a star as being in the RGB or core-He-burning phase, based on the oscillation power spectrum. The network has been trained on power spectra from 82-d time series, corresponding to the K2 time-series for $\epsilon$\,Tau, for which it has a classification accuracy of 95.4 per cent \citep{Hon18}. The neural network classified $\epsilon$\,Tau as a core-He-burning clump star, with a probability of $p=0.986\pm0.012$, where 1 corresponds to a clump star and 0 to a RGB star. Hence, comparing the $\epsilon$\,Tau oscillation spectrum to the oscillation spectra of tens of thousands of red giants observed by the {\it Kepler} telescope, points to the clump as the current evolutionary status of $\epsilon$\,Tau.

\section{Oscillation amplitudes and amplitude ratio}

Apart for the Sun, the amplitude ratio between intensity measurements and radial velocities for solar-like oscillations as a function of frequency has so far only been observed for Procyon \citep{Huber11}. The 
amplitude ratio as a single value has also been derived for $\epsilon$\,Oph \citep{Kallinger08}.
For Procyon, \citet{Huber11} used observations from {\it MOST} \citep{Walker03,Matthews07} that were obtained simultaneously with the RV campaign of \citet{Arentoft08} to derive an amplitude ratio between intensity and RV of 
$A_{\ell=0, Phot}/A_{\ell=0,RV}=0.23\pm0.01$\,ppm\,cm$^{-1}$\,s. Determining the amplitude ratio allows 
tests of scaling relations \citep{Kjeldsen1995,Kjeldsen2011}, and allows
a more robust comparison between observations and models than does comparing the intensity and RV amplitudes individually \citep{Houdek10}.

Amplitudes can be derived following the method of \citet{Kjeldsen08}, which involves heavily smoothing the power spectrum with a Gaussian, converting the spectrum to power density and scaling the signal to {\it amplitude per radial mode}. 
This method was used by \citet{Huber11}, who compared the results for Procyon to the expected amplitude
ratio based on the scaling relations in \citet{Kjeldsen1995}. They found the measured amplitude ratio to be considerably higher than the
predicted value, by about 35\%. They also found the amplitude ratio measured at the frequency of maximum power to be in better agreement with models from \citet{Houdek10} than with the scaling relations. 

The method for deriving the amplitude per radial mode, which involves smoothing the power spectrum with a Gaussian with a full width at half maximum (FWHM) of 4$\Delta\nu$ \citep{Kjeldsen08}, was developed for solar-like main-sequence stars, which have higher oscillation frequencies and broader oscillation envelopes than is the case for evolved stars like $\epsilon$\,Tau. When we applied the width of 4$\Delta\nu$ to $\epsilon$\,Tau, it was evident from the resulting, smoothed power spectra of the SONG and K2 data, that the oscillation signal was too heavily smoothed.
We therefore tested FWHM-values of 2$\Delta\nu$ and 3$\Delta\nu$, and found that those widths worked better for $\epsilon$\,Tau, as the oscillation signal (similar to the white curves in Fig.\,\ref{Fig.PS}) was better retained after smoothing than was the case when using the width of 4$\Delta\nu$. The results for the amplitude ratio of $\epsilon$\,Tau using these two different widths (2$\Delta\nu$ and 3$\Delta\nu$) agreed within the uncertainties. We used the width of 3$\Delta\nu$ in the analysis described below.  

\begin{figure}
   \centering
   \includegraphics[width=9cm]{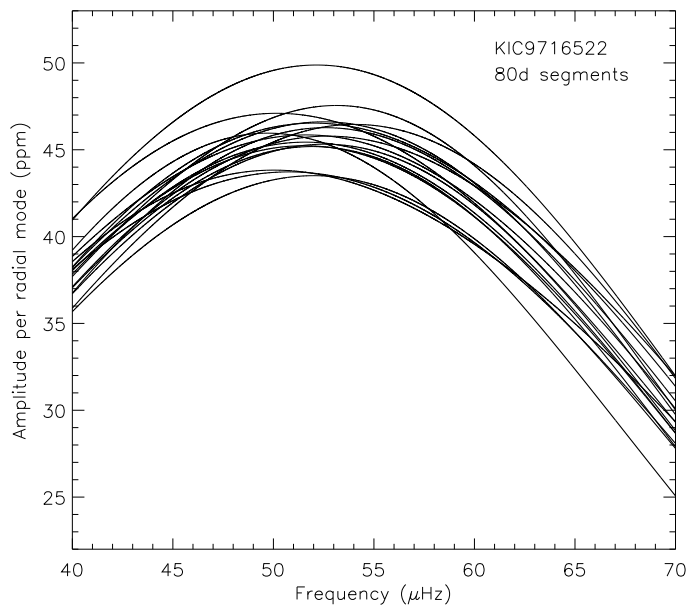}
     \caption{Smoothed power spectra based on 80\,d segments for KIC9716522 in NGC\,6811, which is similar to $\epsilon$\,Tau and which was observed for four years by {\it Kepler}. The data were used to estimate the uncertainty arising from the fact that the SONG and K2 data for $\epsilon$\,Tau were non-simultaneous. 
             }
         \label{Fig.9716522}
   \end{figure}

In the case of $\epsilon$\,Tau, the photometric K2 data and the SONG RV data were not obtained simultaneously.  We can still measure the amplitude ratio but, given the stochastic nature of the 
solar-like oscillations, there will be an additional source of uncertainty in our measurement. In order to quantify this, we used the 18 low-frequency filtered, 80-d segments of the KIC9716522 {\it Kepler}-data described in Sec.~\ref{sec:power-spectrum}. We determined the amplitude per radial mode following \citet{Kjeldsen08} for each of these segments (see Fig.\,\ref{Fig.9716522}), where we smoothed the power spectrum with a Gaussian with a FWHM of 3$\Delta\nu$, and used $c=3.06$ based on a central wavelength of 650\,nm for K2 observations. 
We found a mean amplitude for KIC9716522 of 45.8 ppm, with a standard deviation from the 18 segments of 1.6 ppm, corresponding to an uncertainty of 3.5\%. We have no available data which allow us to perform a similar analysis for RV data, but it seems reasonable to assume that the effect will be similar, and we adopt 4\% as our uncertainty estimate for the photometric and radial velocity amplitudes.

\begin{figure}
   \centering
    \includegraphics[width=9cm]{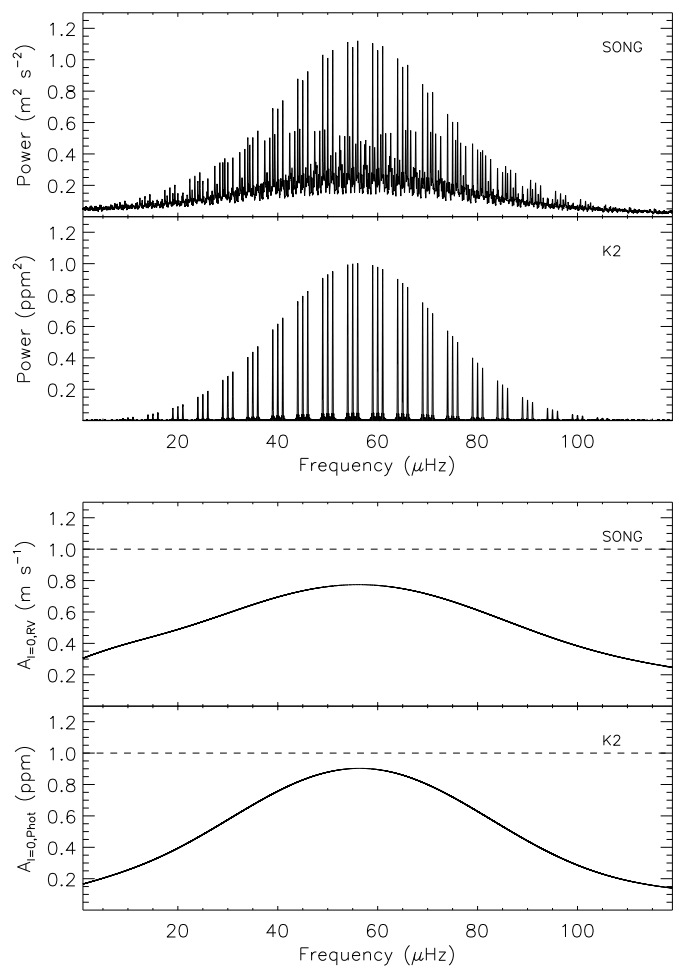}
     \caption{Simulated power spectra to estimate the effect of the window function on the determination of oscillation amplitudes, see text.}
         \label{Fig.simamp}
   \end{figure}

Another factor to consider is the very different spectral windows for the SONG and K2 data (see Fig.\,\ref{Fig.PS}). The K2 spectral window is very clean (i.e. no sidelobes), while the SONG spectral window contains significant sidelobes. Because the distribution of power in the spectral window covers a relatively
broad frequency range compared with the oscillation envelope for $\epsilon$\,Tau, some fraction of the power will be located outside the frequency region
of the oscillations. This will be the case for both the K2- and the SONG-data; however, the effect is expected to be larger for the SONG data
due to the worse spectral window. We would expect to underestimate the oscillation amplitudes due to this effect, and more so for
SONG than for K2. In order to quantify this effect, we performed the simulations illustrated in Fig.\,\ref{Fig.simamp}.
We created a simulated power spectrum based on the oscillation envelope observed for $\epsilon$\,Tau (see Fig.\,\ref{Fig.PS}). Using the 
observed $\Delta\nu$ of 5.0\,$\mu$Hz, we placed a number of oscillation orders between 10 and 110\,$\mu$Hz, and placed 3 peaks in each 
order, so that the factor $c$ \citep{Kjeldsen08} was set at $c=3$. The heights (in power) were distributed according to the observed oscillation envelope, peaking at 1.0 at $\nu_{\rm max}$. We convolved this synthetic spectrum with the SONG and the K2
spectral windows, resulting in the spectra shown in the two upper panels of Fig.\,\ref{Fig.simamp}. We then 
followed the prescription for determining the amplitude per radial mode from \citet{Kjeldsen08}, both using 2$\Delta\nu$ and 3$\Delta\nu$
for the width in the Gaussian smoothing. 

The results are shown in the two bottom panels of Fig.\,\ref{Fig.simamp} for a width of 3$\Delta\nu$. If there were no effect of the spectral window function on the derived amplitudes, the maximum amplitude in both curves would reach 1.0. This is not the case and, as expected, the simulated SONG spectrum is more affected than the simulated K2 spectrum: the amplitude per radial mode reaches a maximum of 0.774\,m\,s$^{-1}$ for SONG, and 0.903\,ppm for K2. This means that we will
underestimate the amplitudes of $\epsilon$\,Tau by similar factors and hence that we will underestimate the RV amplitude more than the
intensity amplitude. We therefore used these two numbers to correct the amplitudes determined from the SONG and the K2 data.

   \begin{figure}
   \centering
   \includegraphics[width=9cm]{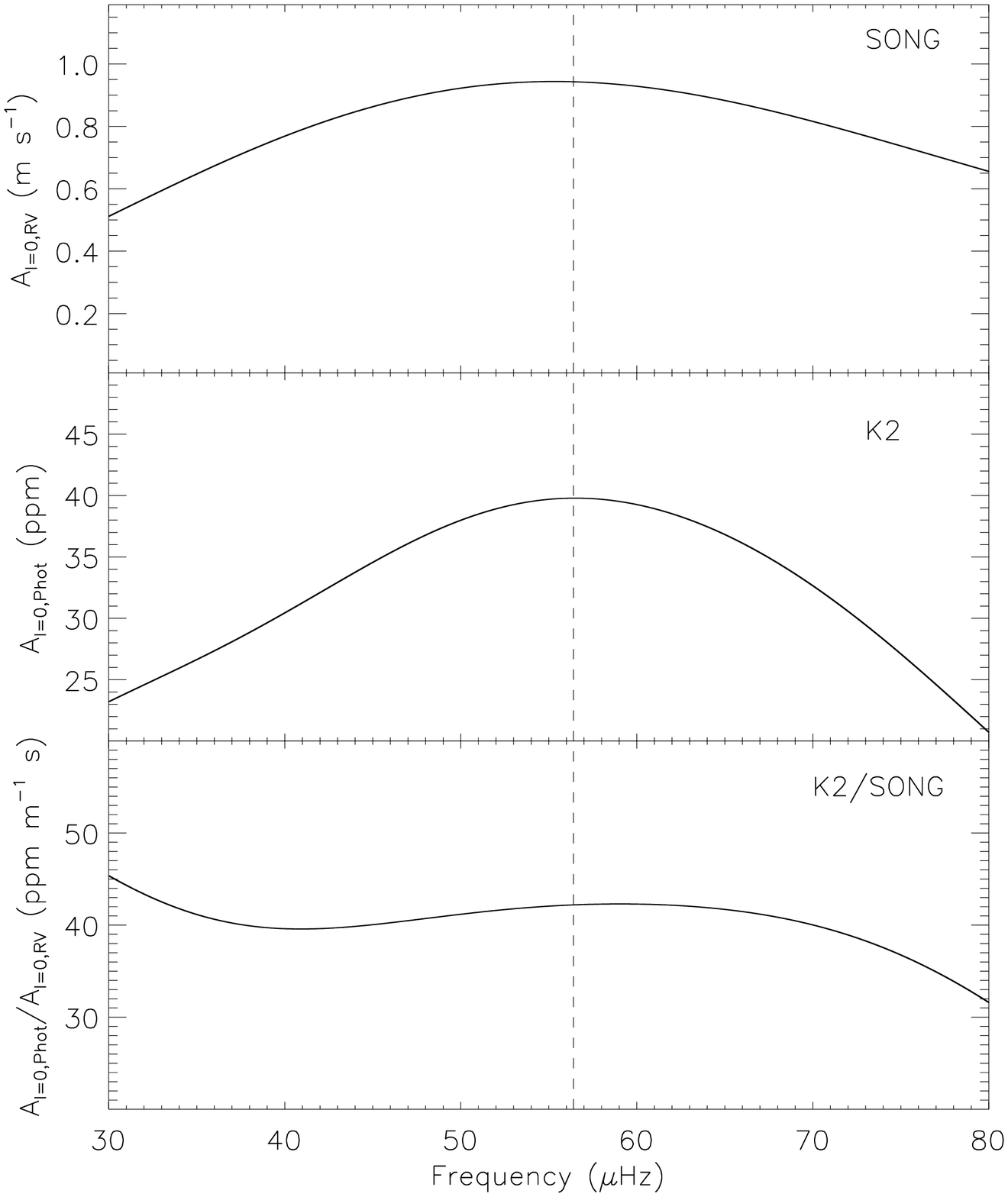}
     \caption{Measured amplitudes in RV (top panel) and intensity (middle panel), and the amplitude ratio as a function of frequency (bottom panel), for $\epsilon$\,Tau. The dashed line indicates $\nu_{\rm max}$}
         \label{Fig.amprat}
   \end{figure}

The final step was to determine the amplitude ratio for $\epsilon$\,Tau. The power spectra were again smoothed using a Gaussian with a FWHM of 3$\Delta\nu$, converted to power density, and converted to amplitude per radial mode using $c=4.09$ for the SONG radial velocities and $c=3.06$ for the K2 data. Using the corrections described above, we measured the amplitudes of $\epsilon$\,Tau to be $A_{\ell=0,RV}=0.94\pm0.04$\,m\,s$^{-1}$ and $A_{\ell=0, Phot}=39.8\pm1.4$\,ppm (see Fig.\,\ref{Fig.amprat}), and hence estimated the amplitude ratio between intensity and RV to be $A_{\ell=0, Phot}/A_{\ell=0,RV}=42.2\pm2.3$\,ppm\,m$^{-1}$\,s (lower panel in Fig.\,\ref{Fig.amprat}). We note that the uncertainty on the amplitude ratio was estimated based on the analysis of the KIC9716522 data described above, and does not include systematic effects arising from, for example, the way we treat the noise background in the data. The true uncertainty may therefore be larger.

The value for the amplitude ratio is significantly higher than expected from scaling relations: using the relations in 
\citet{Kjeldsen1995} and repeated in \citet{Huber11}, we obtain an expected value of only 23.2\,ppm\,m$^{-1}$\,s. We are not able to explain this difference without model calculations, which are beyond the scope of this paper, but we note again that the amplitude ratio of Procyon was 
also higher than expected from the scaling relations \citep{Huber11}. We do not see a shift in our $\epsilon$\,Tau data towards higher frequencies for $\nu_{\rm max}$ in intensity as compared to velocity; such a shift was observed for Procyon \citep{Huber11}. 
Finally, the shape of the amplitude ratio as a function of frequency for $\epsilon$\,Tau agrees better with the models of \citet{Houdek10} than was the case for Procyon, see \citet{Huber11}, their Fig.\,9. We note that the rise in the amplitude ratio at frequencies below 40\,$\mu$Hz in the lower panel in Fig.\,\ref{Fig.amprat} is due to the residual granulation noise at low frequencies in the K2 power density spectrum which is visible in Fig.\,\ref{Fig.PS} (compare the upper and lower panels of Fig.\,\ref{Fig.PS} at frequencies below 40\,$\mu$Hz), so this part of the amplitude ratio curve should be discarded in a forthcoming comparison with model calculations. 

The intensity and RV measurements of $\epsilon$\,Tau were not obtained simultaneously, which weakens conclusions based on the measured amplitude ratio. However, it does seem that the amplitude ratio is higher than expected from simple scaling relations and in better agreement with the model calculations of \citet{Houdek10}. Our results act as a proof-of-concept for combining RV measurements from SONG with space-based intensity data for measuring amplitude ratios of solar-like stars. This opens for new opportunities with the recent launch of {\it TESS} \citep{Ricker15}, which will observe bright, solar-like stars for which we will obtain simultaneous SONG RV time-series data. 

\section{Updated parameters for the planetary system}

\citet{Sato2007} found a planetary companion to $\epsilon$\,Tau, which was the first exoplanet found in an open cluster. Based on a stellar mass of $M=2.7\pm0.1\, \rm M_{\odot}$, they derived a planetary mass of $m_{2}\,\rm{sin}\,i = 7.6\pm 0.2\, \rm M_{\rm J}$. With our slightly lower stellar mass of $M=2.458\pm0.073\, \rm M_{\odot}$,  in agreement with \citet{stello2017}, we can redetermine the minimum mass of the planet, using the velocity semi-amplitude, orbital period, and eccentricity presented in \citet{Sato2007}. We find the new minimum mass of the planet to be $m_{2}\,\rm{sin}\,i = 7.1\pm 0.2\, \rm M_{\rm J}$. 

\section{Conclusions and outlook}

We have determined asteroseismic parameters and combined those with interferometric and spectroscopic data to derive physical parameters for $\epsilon$\,Tau, including its mass.  This leads to a slightly lower revised mass for its planetary companion. 
By combining high-S/N radial-velocity ground-based data from SONG with continuous space-based data from K2, we were able to extract 27 individual oscillation modes in addition to the global asteroseismic parameters.
Although $\epsilon$\,Tau most likely is a (secondary) clump star,
various signatures, including the asteroseismic quantities, model fits and deep learning, gave diverging results on the evolutionary stage of the star, making it unclear whether the star is on the red-giant branch or in the core-helium-burning clump stage.
Due to the fact that $\epsilon$\,Tau displays a relatively sparse set of $\ell$=1 modes, we were unable to determine its asymptotic period spacing, which would allow us to 
determine if the star is in the red-giant-branch phase or already in the core-helium-burning phase \citep{Bedding11}. 
In NGC\,6811, \citet{Arentoft17} found two groupings within the eight secondary clump stars in that cluster; four of the giants displayed rich sets of $\ell$=1 modes, which allowed for a determination of the asymptotic period spacing, while the other four stars were more like $\epsilon$\,Tau, and no period spacing could be determined. If something similar is at play among the Hyades giants, time-series observations of the three other giants may allow for a determination of the asymptotic period spacing for some of the stars, and hence allow us to determine their evolutionary stage.

We have demonstrated 
the potential of combining ground-based data with space-based data, taking advantage of the strengths of both types of observations. Although the spectroscopic and intensity time-series were non-simultaneous, we were also able to determine the amplitude ratio between intensity and radial velocities as a function of frequency, which previously has been done only for the Sun and Procyon. 
The determination of the amplitude ratio of the $\epsilon$\,Tau oscillations was found to be higher than expected from simple scaling relations. This illustrates the importance of obtaining measurements for a larger number of stars, and we plan to do this by obtaining ground-based radial-velocities with SONG that are simultaneous with {\it TESS} observations.
  
\begin{acknowledgements}
Funding for the Stellar Astrophysics Centre is provided by The Danish National Research Foundation (Grant agreement no.: DNRF106). The research was supported by the ASTERISK project
(ASTERoseismic Investigations with {\it SONG} and {\it Kepler}) funded by the European Research Council (Grant agreement no.: 267864).
This work was performed in part under contract with the Jet Propulsion Laboratory (JPL) funded by NASA through the Sagan Fellowship Program executed by the NASA Exoplanet Science Institute.
D.H. acknowledges support by the National Aeronautics and Space Administration under Grants NNX17AF76G and 80NSSC18K0362 issued through the K2 Guest Observer Program. This paper includes data collected by the K2 mission. Funding for the K2 mission is provided by the NASA Science Mission directorate.
\end{acknowledgements}

\bibliographystyle{aa} 
\bibliography{aanda.bib}

\end{document}